\documentclass[aps,prb,twocolumn,showpacs,superscriptaddress,amsmath]{revtex4-1}
\usepackage{graphicx}
\usepackage{bm}
\renewcommand{\vec}[1]{\bm{#1}}

\begin{document}

% Use the \preprint command to place your local institutional report
% number in the upper righthand corner of the title page in preprint mode.
% Multiple \preprint commands are allowed.
% Use the 'preprintnumbers' class option to override journal defaults
% to display numbers if necessary
%\preprint{}
%Title of paper

\title{Magnetic phases of spin-$\frac{3}{2}$ fermions on a spatially anisotropic square lattice}

% repeat the \author .. \affiliation  etc. as needed
% \email, \thanks, \homepage, \altaffiliation all apply to the current
% author. Explanatory text should go in the []'s, actual e-mail
% address or url should go in the {}'s for \email and \homepage.
% Please use the appropriate macro foreach each type of information

% \affiliation command applies to all authors since the last
% \affiliation command. The \affiliation command should follow the
% other information
% \affiliation can be followed by \email, \homepage, \thanks as well.
\author{A. K. Kolezhuk}
%\email[]{Your e-mail address}
%\homepage[]{Your web page}
%\thanks{}
%\altaffiliation{}
\affiliation{Institute of Magnetism, National Academy of Sciences and
  Ministry of Education, 36-b Vernadskii av., 03142 Kiev, Ukraine}
\affiliation{Institute of High Technologies, T. Shevchenko Kiev
  National University, 64 Volodymyrska str., 01601 Kiev, Ukraine}

\author{T. Vekua}
\affiliation{Institut f\"ur Theoretische Physik, Leibniz Universit\"at
  Hannover, Appelstr. 2, 30167 Hannover, Germany}

\date{\today}

\begin{abstract}
We study the magnetic phase diagram of spin-$\frac{3}{2}$ fermions in
a spatially
anisotropic square optical lattice at quarter filling (corresponding
to one particle per lattice site). In the limit of the large on-site
repulsion the system can be mapped to the so-called $Sp(N)$ Heisenberg
spin model with $N=4$. We analyze the $Sp(N)$ spin model with the help
of the large-$N$ field-theoretical approach and show that the
effective theory corresponds to the $Sp(N)$ extension of the
$CP^{N-1}$ model, with the Lorentz invariance  generically
broken. We obtain the renormalization flow of the model couplings and
show that although the $Sp(N)$ terms are seemingly irrelevant, their
presence leads to a renormalization of the $CP^{N-1}$ part of the
action, driving a phase transition. 
We further consider the influence of the external
magnetic field (the quadratic Zeeman effect), and 
present the qualitative
analysis of the ground state phase diagram.

\end{abstract}

% insert suggested PACS numbers in braces on next line

% insert suggested keywords - APS authors don't need to do this
%\keywords{}

%\maketitle must follow title, authors, abstract, \pacs, and \keywords
\maketitle

%%%%%%%%%%%%%%%%%%%%%%%%%%%%%%%%%%%%%%%%%%%%%%%%%%%%%%%%%%%%%%%%%%%%%%%%%%%%%%%%
\section{Introduction}

Extraordinary controllability of ultracold gases allows highly
accurate modeling and study of problems originated in condensed matter
physics. Frustrated magnetic systems occupy an important position on
the list of intriguing problems that could be studied in
multicomponent ultracold gases.  Recently, multicomponent ultracold
Fermi gases have attracted much attention
\cite{Lewenstein+rev07,Bloch+rev08} motivated by the growing
availability of hyperfine-degenerate fermionic atoms, such as
${}^{6}$Li, \cite{Ottenstein+08,Huckans+09,Wenz+09} ${}^{40}$K,
\cite{Modugno+03} ${}^{135}$Ba and ${}^{137}$Ba, \cite{He+91} and
${}^{173}$Yb. \cite{Fukuhara+07} Realization of unconventional phases
of $SU(N)$ internally frustrated antiferromagnets\cite{AffleckMarston}
have been recently suggested\cite{Gorshkov,Hermele} in alkaline earth
atoms with nuclear spin as large as $\frac{9}{2}$ in ${}^{87}$Sr.

Among multicomponent ultracold gases, spin-$\frac{3}{2}$ alkaline
fermions stand out by their rich physics 
characterized by an enlarged $Sp(4)$ symmetry,
which is naturally present in the system without fine-tuning of any
parameters. \cite{WuHuZhang03} By tuning the ratio of scattering lengths in the two
allowed spin-$0$ and spin-$2$ channels, even the larger $SU(4)$ symmetry may
be achieved.\cite{WuHuZhang03,Lecheminant+05,Wu05,Wu06} 

Experiments with ultracold atoms are usually done in the presence of
magnetic fields. For atoms with hyperfine spins $F\geq 1$, the
spin-changing collisions redistribute the populations of the
components with different spin projection $F^{z}$ while keeping the
total magnetization $M=\sum_{j}F^{z}_{j} $ fixed. Therefore, the usual
linear Zeeman effect induced by an external magnetic field does not
play any role for a state with a fixed initially prepared $M$, and the
main influence of an external magnetic field (except the change of
scattering lengths due to the Feshbach resonance phenomenon) is
contained in the quadratic Zeeman effect (QZE). The quadratic Zeeman
field $q$ couples to $\sum_{j}(F_{j}^{z})^{2}$ and thus introduces a
difference in chemical potentials for components with different
$|F^{z}|$.  A peculiar property of spin-$\frac{3}{2}$ fermions is the
fact that even in presence of the quadratic Zeeman field, the high
$Sp(4)$ symmetry is lowered to $SU(2)\times SU(2)$ and thus remains quite
high. \cite{Temo-spin32}

\begin{figure}[b]
 \includegraphics[width=0.36\textwidth]{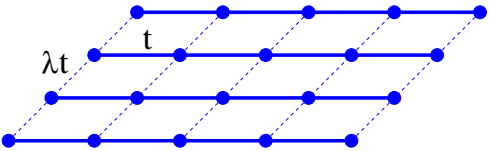}
 \caption{
\label{fig:lattice} Square lattice with the anisotropic hopping considered in this paper.
The hopping amplitudes are $t$ and $\lambda t$  along  the horizontal
and vertical bonds, respectively. }
\end{figure}

In this work, we study the magnetic phase diagram of
spin-$\frac{3}{2}$ fermions at quarter filling, in the limit of a
strong on-site repulsion, on an anisotropic square lattice with
hopping amplitudes in two spatial directions differing by the factor
$\lambda\in[0,1]$, as depicted in Fig.\ \ref{fig:lattice}.  We
construct the effective field theory describing the low-energy
properties of the system which has the form of a $Sp(N)$ extension of
the $CP^{N-1}$ model, with the generically broken Lorentz
invariance. For this field theory, the analysis of the one-loop
renormalization group equations shows that the $Sp(N)$ terms are
dangerously irrelevant: their presence leads to a considerable
renormalization of the $CP^{N-1}$ part of the action.  As the result,
by changing the ratio of the scattering lengths or the lattice
anisotropy parameter $\lambda$ one can drive the phase transition
between the long range ordered N\'eel state and the valence-bond-solid
(VBS) state.

We also include into consideration the quadratic Zeeman coupling and
study the evolution of the ground state under QZE.  Since the QZE
preserves the $SU(2)$ symmetry,\cite{Temo-spin32} the ground state at
large values of the quadratic Zeeman field $q$ corresponds to the
long-range ordered (N\'eel) phase of the isotropic spin-1/2 Heisenberg
antiferromagnet (HAFM), for any nonzero value of the lattice
anisotropy parameter $\lambda \neq 0$.  We show that, depending on
the value of the anisotropy $\lambda$, when the field $q$ is
decreased, this state either adiabatically evolves into the N\'eel
phase of 4-component fermions or
undergoes a phase transition into the VBS state.

The structure of the paper is as follows: in
Sect.\ \ref{sec:efftheory} we present the derivation of the low-energy
effective field theory for a system spin-$\frac{3}{2}$ fermions at quarter
filling in the regime of a Mott insulator (in other words, in the
regime of the $Sp(4)$ Heisenberg model). In Sect.\ \ref{sec:RG} we
analyze the renormalization group flow of the derived model and sketch
the phase diagram of the system in dimensions one and
two. In Sect.\ \ref{sec:Q} we study the  effect of an external
quadratic Zeeman field; finally, Sect.\ \ref{sec:summary} contains the
summary and discussion of the results.

%%%%%%%%%%%%%%%%%%%%%%%%%%%%%%%%%%%%%%%%%%%%%%%%%%%%%%%%%%%%%%%%%%%%%%%%%%%%%%%%
\section{\boldmath Effective low-energy field theory for the $Sp(N)$ Heisenberg model}
\label{sec:efftheory}
%%%%%%%%%%%%%%%%%%%%%%%%%%%%%%%%%%%%%%%%%%%%%%%%%%%%%%%%%%%%%%%%%%%%%%%%%%%%%%%%

Consider a system of spin-$\frac{3}{2}$ fermions on  a
two-dimensional anisotropic square lattice. In the $s$-wave scattering approximation,
this system can be described by the following Hamiltonian: \cite{WuHuZhang03}
\begin{eqnarray} 
\label{ham-atom}
\widehat{H}&=&-\sum_{\sigma=\pm 1/2,\pm 3/2}\sum_{\langle
  ij\rangle}t_{ij}(c^{\dag}_{\sigma,i}c^{\vphantom{\dag}}_{\sigma,j} +\text{h.c.})
\nonumber\\ 
&+&\sum_{i}\sum_{F=0,2}U_{F}\sum_{m=-F}^{F} P^{\dag}_{Fm,i}  P^{\vphantom{\dag}}_{Fm,i} ,
\end{eqnarray}
where $c_{\sigma,i}$ are the spin-$\frac{3}{2}$ fermionic operators at
the lattice site $i$, $t_{ij}$ are the effective hopping amplitudes between
two neighboring sites, 
 $P_{Fm,i}=\sum_{\sigma\sigma'}\langle
Fm|\frac{3}{2} \sigma,\frac{3}{2} \sigma' \rangle
c_{\sigma,i}c_{\sigma',i}$ are the operators describing an on-site
pair with the total spin $F$, and the positive interaction constants
$U_{0}$, $U_{2}$ are proportional to the scattering lengths in the
$F=0$ and $F=2$ channels, respectively. The hopping  is
assumed to be generally anisotropic in two spatial directions, i.e.,
\begin{equation} 
\label{hop}
t_{ij}=
\begin{cases} t & \text{for $(ij)\parallel \widehat{\vec{x}} $}\\
\lambda t & \text{for $(ij)\parallel \widehat{\vec{y}}$} 
\end{cases},\quad
0<\lambda\leq 1.
\end{equation}
Although our main interest will be in the behavior of the two-dimensional model, we will also make a few
comments about the one-dimensional case which formally corresponds to $\lambda=0$.

We will be also interested in the effect of the external magnetic
field. Since the total magnetization in cold atom experiments has very
long relaxation time, the primary effect of the external field is
given by the quadratic Zeeman term:
\begin{equation} 
\label{q-Zeeman}
\widehat{H}_{Z}=q\sum_{i\sigma}\sigma^{2} c^{\dag}_{\sigma,i}c^{\vphantom{\dag}}_{\sigma,i}.
\end{equation}

At quarter filling (one particle per site), and in the limit of strong
on-site repulsion $t_{ij} \ll U_{0},U_{2}$ the charge degrees of freedom
are strongly gapped and the system can be approximately described by
an effective spin Hamiltonian, which can be
conveniently written in the following form:\cite{WuHuZhang03,Lecheminant+05,Wu05,Wu06}
\begin{eqnarray} 
\label{ham-spin} 
&& \widehat{H}_{\rm spin}= \sum_{\stackrel{1\leq a< b\leq 5}{\langle ij\rangle}} 
J_{1,ij}\widehat{\Gamma}^{ab}_{i}\widehat{\Gamma}^{ab}_{j}
-\sum_{\stackrel{1\leq a \leq 5}{\langle ij\rangle}} 
J_{2,ij}\widehat{\Gamma}^{a}_{i}\widehat{\Gamma}^{a}_{j}, \\
&& \widehat{\Gamma}^{ab}_{i}=c^{\dag}_{\sigma,i}
\Gamma^{ab}_{\sigma\sigma'}c_{\sigma',i},\qquad
\widehat{\Gamma}^{a}_{i}=c^{\dag}_{\sigma,i}
\Gamma^{a}_{\sigma\sigma'}c_{\sigma',i},\nonumber
\end{eqnarray}
where  the $4\times4$ matrices $\Gamma^{a}$ and
$\Gamma^{ab}=\frac{1}{2i}[\Gamma^{a},\Gamma^{b}]$  are
the generators of the $SU(4)$ Lie algebra. 
The operators
$\widehat{\Gamma}^{a}$ form a vector representation of the $Sp(4)$
group, and $\widehat{\Gamma}^{ab}$  form an adjoint representation of
$Sp(4)$, so the Hamiltonian (\ref{ham-spin}) is explicitly $Sp(4)$-invariant. 

In the second order of the perturbation theory in $t$, the exchange constants
$J_{1,2}$ are given by:
\begin{equation} 
\label{J12} 
 J_{1,ij}=\frac{1}{4}\Big(\frac{t_{ij}^{2}}{U_{0}}+\frac{t_{ij}^{2}}{U_{2}} \Big)
,\quad
 J_{2,ij}=\frac{1}{4}\Big(\frac{t_{ij}^{2}}{U_{0}}-\frac{3t_{ij}^{2}}{U_{2}}
 \Big) .
\end{equation}
The operators $\widehat{\Gamma}^{a}$, $\widehat{\Gamma}^{ab}$ can be
expressed in terms of four Schwinger bosons $b_{\alpha}$,
$1\leq \alpha\leq 4$, satisfying the constraint
$b^{\dag}_{\alpha}b_{\alpha}=1$
at each site: effectively one can
just replace the operators $c_{\sigma,i}$ by $b_{\alpha,i}$ in the
definition of $\widehat{\Gamma}$ in (\ref{ham-spin}). Doing so, one
arrives at the Hamiltonian of the form\cite{QiXu08}
\begin{eqnarray} 
\label{ham-b} 
\widehat{H}_{\rm spin} &=&  \sum_{\langle ij\rangle}
J'_{ij} K^{\dag}_{ij}K_{ij} -J_{ij} Q^{\dag}_{ij}Q_{ij},\\
&& K_{ij}=b^{\dag}_{\alpha,i}b_{\alpha,j},\quad Q_{ij}=b_{\alpha,i}\mathcal{J}_{\alpha\beta}b_{\beta,j},\nonumber
\end{eqnarray}
where $\mathcal{J}$ is the antisymmetric matrix that has the
following properties: 
\begin{eqnarray*} 
\label{Jantisym} 
&& \mathcal{J}^{T}=\mathcal{J}^{\dag}=-\mathcal{J},
\mathcal{J}^{2}=-1,\\
&& \mathcal{J}^{\dag}\Gamma^{a}\mathcal{J}=(\Gamma^{a})^{T}, \quad
\mathcal{J}^{\dag}\Gamma^{ab} \mathcal{J}=-(\Gamma^{ab})^{T}. 
\end{eqnarray*}
One specific representation of the matrices introduced above is given
by: \cite{QiXu08}
\begin{eqnarray} 
\label{matrices} 
&&\Gamma^{a}=\sigma^{a}\otimes \sigma^{z},\; a=1,2,3,\quad
\Gamma^{4}=1\otimes\sigma^{x},\nonumber\\
&& \Gamma^{5}=1\otimes \sigma^{y},\quad
\mathcal{J}=i\sigma^{y}\otimes \sigma^{x},
\end{eqnarray}
where $\sigma^{a}$ are the Pauli matrices.  The $Sp(4)$ group may be
viewed as a group of unitary $4\times4$ matrices $U$ that satisfy the
condition $U^{T}\mathcal{J} U = \mathcal{J}$.

The couplings $J$, $J'$ in (\ref{ham-b}) are given by:
\begin{equation} 
\label{JJ'} 
J_{ij}=\frac{t_{ij}^{2}(U_{2}-U_{0})}{U_{0}U_{2}},\quad 
J'_{ij}=\frac{2t_{ij}^{2}}{U_{2}}
\end{equation}
and are assumed to be positive; in terms of the atomic
spin-$\frac{3}{2}$ system this corresponds to the assumption
\[
U_{2}>U_{0}>0.
\]
The Hamiltonian in the form (\ref{ham-b}) can be easily generalized to
the case of an \emph{even} number $N$ of bosonic flavors $b_{\alpha}$,
$\alpha\in[1,N]$ and the local hardcore constraint for the Schwinger
bosons is generalized by
\[
b^{\dag}_{\alpha}b_{\alpha}=n_{c},
\]
with the number $n_{c}$ playing the role of the ``spin magnitude''.\cite{ReadSachdev90}

The $Sp(N)$ symmetry of the Hamiltonian is enlarged to $SU(N)$ at the
point $J=0$, where the two-site Hamiltonian becomes a permutation
operator of two $N$-component objects.  Since the lattice is
bipartite, the enlargement of symmetry to $SU(N)$ happens also at
$J'=0$ point.\cite{WuHuZhang03,Lecheminant+05,Wu05,Wu06,QiXu08} The latter point
$J'=0$ corresponds to a $SU(N)$ antiferromagnet where spins transform
according to the fundamental representations of $SU(N)$ on sublattice
A and according to the conjugate representation on sublattice B. In
the following we will refer to this point as the staggered $SU(N)$
antiferromagnetic point. The other $SU(N)$ point $J=0$, where spins
are in the fundamental representations of $SU(N)$ on each site,
corresponds to the exactly solvable Uimin-Lai-Sutherland model in one
dimension\cite{Uimin70Lai74Sutherland75} and we will call it the
uniform $SU(N)$ antiferromagnetic point.

Our strategy will be to use the staggered $SU(N)$ antiferromagnetic point $J'=0$
(i.e., $U_{2}\to\infty$) as
the starting point to construct the effective low-energy
field theory. 
First of all, we make a unitary transformation\cite{Wu06,QiXu08} on one
sublattice,
\begin{equation} 
\label{unitary} 
\mathcal{S}_{n}\mapsto \mathcal{J}^{\dag} \mathcal{S}_{n} \mathcal{J},
\quad \mathcal{S}\in\{ \widehat{\Gamma}^{a},\widehat{\Gamma}^{ab} \},\quad
n\in\text{sublattice $B$},
\end{equation}
that effectively interchanges the operators $\widehat{Q}$ and
$\widehat{K}$ in the Hamiltonian (\ref{ham-b}). 
Further, we use the usual coherent state path integral
formalism, passing from  the bosonic operators
$\widehat{b}_{\alpha,n}$ to the corresponding $c$-number lattice variables $\vec{b}_{n}$.
It is easy to see that at the mean-field level both terms in the
Hamiltonian (\ref{ham-b}) are \emph{simultaneously} minimized for a uniform distribution of
$\vec{b}_{n}$, provided that
$J,J'>0$. Thus, in the parameter region $J,J'>0$ in (\ref{ham-b}) 
one may expect the physics to be
rather different from that found in $Sp(N)$ models describing
geometrically frustrated systems.\cite{ReadSachdev-spn-91}

In terms of $\vec{b}_{n}$, the Euclidean action on the lattice takes
the form $\mathcal{A}_{\rm lat}=\int d\tau \mathcal{L}_{\rm lat}$,
with the Lagrangian:
\begin{eqnarray} 
\label{Llat} 
\mathcal{L}_{\rm lat}&=&\sum_{n}\eta_{n}\vec{b}_{n}^{*}\cdot
\partial_{\tau} \vec{b}_{n} \nonumber\\
&+&\sum_{\langle nn'\rangle}\Big( 
J'|\vec{b}_{n}\cdot \mathcal{J}\vec{b}_{n'}|^{2} 
-J|\vec{b}_{n}^{*}\cdot \vec{b}_{n'}|^{2}
\Big) .
\end{eqnarray}
In a standard manner, we then split the field $\vec{b}$ into the smooth
and staggered components $\vec{z}_{n}$, $\vec{\psi}_{n}$ :
\begin{equation} 
\label{smooth} 
 \vec{b}_{n} =  \sqrt{n_{c}}(\vec{z}_{n} + \eta_{n}\vec{\psi}_{n}),
\end{equation}
where $\eta_{n}$ takes values of $\pm1$ on $A$ and $B$ sublattices,
respectively, and the constraints
\begin{equation} 
\label{constr} 
|\vec{z}|^{2}+|\vec{\psi}|^{2}=1,\quad    
\vec{\psi}\cdot \vec{z}^{*}+\vec{z}\cdot\vec{\psi}^{*}=0
\end{equation}
are implied.
One can expect that the magnitude of the  staggered component $\vec{\psi}$ that corresponds
to ferromagnetic fluctuations, will be much smaller than that of
$\vec{z}$.

The unitary transformation (\ref{unitary}) is a necessary step: for
positive $J'$ one cannot start from the uniform $SU(N)$ antiferromagnetic
point $J=0$, since no reasonable choice of smooth fields is possible.
It has to be remarked, however, that our choice of smooth fields
becomes poor in the vicinity of the 
point $J=0$ that has a much higher degeneracy of the mean-field ground
state. One may thus expect that the resulting effective field theory
is not reliable close to the uniform $SU(N)$ antiferromagnetic point.

Passing to the continuum and making the gradient expansion,
while retaining only up to quadratic terms in $\vec{\psi}$ and
neglecting its derivatives, one readily obtains the Euclidean action
in the form $\mathcal{A}= \mathcal{A}_{0}+ \mathcal{A}_{\rm int} +
\mathcal{A}_{\rm top}$, where $\mathcal{A}_{0}$ corresponds to $J'=0$:
\begin{eqnarray} 
\label{A0}
\mathcal{A}_{0}&=& \sqrt{\lambda}n_{c}^{2}\int d\tau \int d^{d}x \Big\{  
\frac{1}{n_{c}}(\vec{\psi}^{*}\cdot
\partial_{\tau}\vec{z}-\vec{\psi}\cdot\partial_{\tau}\vec{z}^{*}) \nonumber\\
&+&J \big( |\partial_{k}\vec{z}|^{2}
-|\vec{z}^{*}\cdot\partial_{k}\vec{z}|^{2}\big)\nonumber\\
& +&4J(d-1+\lambda)
\big(|\vec{\psi}|^{2}-|\vec{\psi}^{*}\cdot\vec{z}|^{2}\big) \nonumber\\
&+&\mu_{1}(\vec{\psi}\cdot\vec{z}^{*}+\vec{\psi}^{*}\cdot\vec{z})
+\mu_{2}(|\vec{z}|^{2}+|\vec{\psi}|^{2}-1)
\Big\}.
\end{eqnarray}
The term $\mathcal{A}_{\rm top}$ contains the topological phase
contribution, \cite{Haldane88,ReadSachdev89,ReadSachdev90} 
\begin{equation}
\label{Atop}
\mathcal{A}_{\rm top}[\vec{z}]=n_{c}\int d\tau \sum_{n}
\eta_{n}\vec{z}^{*}_{n}\cdot\partial_{\tau}\vec{z}_{n},
\end{equation}
and the perturbation  $\mathcal{A}_{\rm int}$ is
determined by the deviation $J'$ from the staggered antiferromagnetic $SU(4)$ point:
\begin{equation} 
\label{Aint}
\mathcal{A}_{\rm int}=J'n_{c}^{2}\sqrt{\lambda}\int d\tau \int d^{d}x \Big\{
|\widetilde{\vec{z}}\cdot \partial_{k}\vec{z}|^{2} 
+4(d-1+ \lambda)|\vec{\psi}\cdot\widetilde{\vec{z}}|^{2}
\Big\}.
\end{equation}
Here and throughout the paper we use the notation
\begin{equation} 
\label{ztilde}
\widetilde{\vec{z}}\equiv\mathcal{J}\vec{z} .
\end{equation}
In the above expressions,  $d=1$ or $2$ is the spatial dimension, the index
$k=1,\ldots, d$ labels spatial coordinates, and the lattice constant $a$ and the
Planck constant has been set to unity for convenience.
The factor $\sqrt{\lambda}$  comes from rescaling one of the
coordinates to compensate for the anisotropy of interactions (for $d=1$
one has to set effectively $\lambda=1$), and
$\mu_{1,2}$ are the Lagrange multipliers ensuring the constraints.

Integrating out the staggered component $\vec{\psi}$ can be easily
performed (see Appendix \ref{app:micro} for details), and one arrives at the
following effective action for the  complex unit vector field $\vec{z}$:
\begin{eqnarray} 
\label{Aeff} 
\mathcal{A}_{\rm eff}[\vec{z}]&=&\frac{\Lambda^{d-1}}{2}\int d^{d+1}x \Big\{
\frac{1}{g_{1}}|\mathcal{D}_{0}\vec{z}|^{2}
+\frac{1}{g_{2}}|\mathcal{D}_{k}\vec{z}|^{2} \nonumber\\
&+&\frac{1}{\widetilde{g_{1}}}|\widetilde{\vec{z}}\cdot \mathcal{D}_{0}\vec{z}|^{2} 
+\frac{1}{\widetilde{g_{2}}}|\widetilde{\vec{z}}\cdot \mathcal{D}_{k}\vec{z}|^{2} 
\Big\}+\mathcal{A}_{\rm top}.
\end{eqnarray}
Here $\Lambda$ is the ultraviolet momentum cutoff,
$x^{0}=2J\tau\sqrt{d-1+\lambda}$ is the rescaled imaginary time coordinate,
$\mathcal{D}_{\mu}=\partial_{\mu}-i\mathcal{A}_{\mu}$ is the gauge covariant derivative, and
$\mathcal{A}_{\mu}=-i(\vec{z}^{*}\cdot \partial_{\mu}\vec{z})$ is the gauge
field. 
It is worth
noting that $\widetilde{\vec{z}}\cdot \mathcal{D}_{\mu}\vec{z}\equiv
\widetilde{\vec{z}}\cdot \partial_{\mu}\vec{z}$ since
$\widetilde{\vec{z}}\cdot \vec{z}=0$. 
The bare values of the coupling constants in (\ref{Aeff}) are given
by the following expressions:
\begin{eqnarray} 
\label{bare} 
&& g_{1}^{(0)}=g_{2}^{(0)}=
\frac{1}{n_{c}}\Big\{1+\frac{d-1}{\lambda}\Big\}^{1/2},\nonumber \\
&& \widetilde{g_{1}}^{(0)}=-(1+J/J')g_{1}^{(0)}, 
\quad  \widetilde{g_{2}}^{(0)}=(J/J')g_{2}^{(0)} .
\end{eqnarray}
In particular, note the different signs of $\widetilde{g_{1}}$ and $\widetilde{g_{2}}$.
The first two terms in the action (\ref{Aeff}) constitute  the
familiar action of the 
$CP^{N-1}$
model,\cite{Eichenherr78,GoloPerelomov78,DAdda+78,Witten79,ArefevaAzakov80}
that has been extensively studied as an effective
theory for $SU(N)$ antiferromagnets \cite{ReadSachdev89,ReadSachdev90}. The third and fourth terms
are proportional to $J'$ and thus 
represent perturbation caused by the deviation from the $SU(N)$
staggered antiferromagnetic point. The presence of those terms has
been noticed by Qi and Xu,\cite{QiXu08} but they have been neglected
since they seem to be irrelevant. In the next section, we will show
that those $Sp(N)$ terms are generally only marginally irrelevant, and
can drive a phase transition.

%%%%%%%%%%%%%%%%%%%%%%%%%%%%%%%%%%%%%%%%%%%%%%%%%%%%%%%
Here a remark is in order:  in the action above, for the 2d case we have
effectively removed the lattice anisotropy by rescaling one of the
coordinates. Due to the $Sp(4)$ symmetry of the problem, the remaining perturbations that break the
90 degree rotation symmetry of the lattice appear only in
higher orders in field derivatives: the lowest-order terms of this
type have the form
$f_{\mu}|\vec{z}^{*}\mathcal{D}^{2}_{\mu}\vec{z}|^{2}$ and
$f_{\mu}(\widetilde{\vec{z}}\mathcal{D}_{\mu}\vec{z})(\widetilde{\vec{z}}^{*}D^{2}_{\mu}\vec{z}^{*})+\text{c.c.}$,
with $f_{x}\not= f_{y}$. Such terms are less relevant than those taken
into account in the action (\ref{Aeff}) and thus will be neglected.
%%%%%%%%%%%%%%%%%%%%%%%%%%%%%%%%%%%%%%%%%%%%%%%%%%%%

The properties of the $CP^{N-1}$ model are well understood: in the absence
of the topological term $\mathcal{A}_{\rm top}$ given by (\ref{Atop}), it is always disordered in
$d=1$, and in $d=2$ the long range order appears below a certain critical value
of the effective coupling.\cite{DAdda+78,Witten79} This critical value
depends on $N$, and from the numerical work \cite{Assaad05,KawashimaTanabe07,Beach+09} it is known that
the two-dimensional $CP^{N-1}$ model on a square lattice is
disordered for $N/n_{c}\geq 5$.

In the disordered phase the field $\vec{z}$ acquires a finite mass, and 
a kinetic term for
the gauge field is dynamically generated. \cite{Witten79}
It is also well known that
the topological term becomes crucially important in the disordered
phase; \cite{Haldane88,ReadSachdev89,ReadSachdev90} particularly, it
leads to a
spontaneous dimerization in $d=1$ for odd $n_{c}$ (except for $N=2$ which is
special: in that case the system remains gapless and translationally invariant
in a wide $g$ range\cite{Haldane83,ShankarRead90,Affleck91-prl,Azcoiti+07}), and
in $d=2$ the disordered phase gets spontaneously dimerized in different
patterns depending on the value of $(n_{c} \mod 4)$. The
``disordered'' phase thus  acquires valence bond solid (VBS)
order connected to the broken translational invariance. For the
$Sp(4)$ case, parent Hamiltonians with exact ground states of the VBS
type have been constructed recently.\cite{SchurichtRachel08}

An effective theory for the $Sp(N)$ Heisenberg model in a form similar
to (\ref{Aeff}) has been obtained by Kataoka et al.\cite{Kataoka+10}
However, our result differs from that of Ref.\ \onlinecite{Kataoka+10}
in one important respect: the last two terms in (\ref{Aeff}),
proportional to the ``perturbation'' $J'$, explicitly break the
Lorentz invariance, while in the theory of
Ref.\ \onlinecite{Kataoka+10} the Lorentz invariance is retained even
in the presence of the ``perturbation''.  By a simple classical linear
excitation analysis of the initial lattice action (\ref{Llat}) one can
obtain $(N-1)$ Goldstone modes (``spin waves''), having the velocities
$v_{1}=v_{2}=\ldots =v_{N-2}=1$ and $v_{N-1}=1+J'/J$ (in units of
$2J\sqrt{d}$), see Appendix \ref{app:sw} for details.  Our effective theory
(\ref{Aeff}) yields $(N-2)$ modes with the velocity
$u_{1}=\sqrt{\gamma}$ and one mode with the velocity
$u_{2}=u_{1}\sqrt{(1+\rho)/(1-\kappa)}$, where $\gamma=g_{1}/g_{2}$,
$\rho=g_{2}/\widetilde{g_{2}}$, and
$\kappa=-g_{1}/\widetilde{g_{1}}$. After substituting the bare values
of Eq.\ (\ref{bare}), this is in a perfect agreement with the spin
wave calculation, while the theory of Ref.\ \onlinecite{Kataoka+10}
yields the same velocity $v=1$ for all three modes.

For $N=4$, the contribution of the quadratic Zeeman field (\ref{q-Zeeman}) takes
the following form: 
\begin{equation} 
\label{A-Zeeman} 
\mathcal{A}_{Z}=\frac{\Lambda^{d+1}}{2}m_{0}^{2}\int d\tau \int d^{d}x(|z_{1}|^{2}+|z_{4}|^{2}),
\end{equation}
with the bare ``mass'' value given by
\begin{equation} 
\label{m0} 
m_{0}^{2}=2q/(Jg_{1}^{(0)}).
\end{equation}
For any finite $q$, the symmetry of the model is reduced from $Sp(4)$ to
$SU(2)\times SU(2)$\cite{Temo-spin32}. If $q$ is large enough, the
effective theory reduces to that of a two-component complex field,
i.e., to the $CP^{1}$ model.

%%%%%%%%%%%%%%%%%%%%%%%%%%%%%%%%%%%%%%%%%%%%%%%%%%%%%%%%%%%%%%%%%%%%%%%%%%%%%%%%
\section{One-loop renormalization group analysis at zero field}
\label{sec:RG}
%%%%%%%%%%%%%%%%%%%%%%%%%%%%%%%%%%%%%%%%%%%%%%%%%%%%%%%%%%%%%%%%%%%%%%%%%%%%%%%%

Consider first the case when the external field is absent.
To understand the role of the $Sp(N)$ terms in the action (\ref{Aeff}),
we have to analyze their behavior under renormalization.
Renormalization group (RG) equations for  spin liquids described
by a Lorentz-invariant
low-energy field theory with $SU(N)$ and $Sp(N)$
symmetries have been studied recently\cite{CenkeXu2008} by means of
the fermionic large-$N$  formulation.
 Our
effective theory (\ref{Aeff}) does not possess the
Lorentz invariance.
 We write down
one-loop RG equations for the model (\ref{Aeff}), using Polyakov's background
field  method.\cite{Polyakov75} The details of the derivation can be found in Appendix
\ref{app:rg}. It is convenient to define the following parameters:
\begin{equation} 
\label{rg-pars} 
\gamma=\frac{g_{1}}{g_{2}},\quad y=2C_{d}\sqrt{g_{1}g_{2}}, \quad
\kappa=-\frac{g_{1}}{\widetilde{g_{1}}},\quad \rho=\frac{g_{2}}{\widetilde{g_{2}}},
\end{equation}
where $C_{d}=\pi S_{d}/(2\pi)^{d+1}$ and
$S_{d}=2\pi^{d/2}/\Gamma(d/2)$ is the surface of a $d$-dimensional sphere.
The minus sign has been introduced in $\kappa$, to compensate
for the negative initial value of $\widetilde{g_{1}}$ in
(\ref{bare}). The physical meaning of the couplings (\ref{rg-pars})
will be clarified below.
Their bare values are:
\begin{eqnarray} 
\label{bare-rg} 
&&
y^{(0)}=2C_{d}g_{1}^{(0)},\quad
\gamma^{(0)}=1, \nonumber \\
&& \kappa^{(0)}=\frac{J'}{J+J'}, 
\quad  \rho^{(0)}=\frac{J'}{J} .
\end{eqnarray}
The resulting RG equations can be conveniently written
down in the following form:
\begin{eqnarray} 
\label{rg-eqs} 
\dot{y}&=&(1-d)y+y^{2}\Big\{N-1\nonumber\\
&+&\frac{1}{2}(\kappa-\rho)
+\frac{1}{[(1-\kappa)(1+\rho)]^{1/2}}
\Big\},\nonumber\\
\dot{\gamma}&=&y\gamma(\kappa+\rho),\\
\dot{\kappa}&=& y \left\{
\kappa^{2}-(N-3)\kappa -1 +\Big(\frac{1-\kappa}{1+\rho}\Big)^{1/2}
\right\},\nonumber\\
\dot{\rho}&=& y\left\{ 
1-\rho^{2}-(N-3)\rho -\Big(\frac{1+\rho}{1-\kappa}\Big)^{1/2}
\right\},\nonumber
\end{eqnarray}
where a dot denotes the derivative $d/dl = -\Lambda(d/d\Lambda)$ with respect to the scale variable.

For the staggered antiferromagnetic $SU(N)$ point $J'=0$ we have
$\kappa^{(0)}=0$, $\rho^{(0)}=0$, and the above equations reduce to the single
equation for the coupling $g=y/(2C_{d})$ of the $CP^{N-1}$ model
in $d$ spatial dimensions. For $d=2$ this model is ordered ($g$
renormalizes to zero), if the ratio $n_{c}/N$ is above a certain
critical value;\cite{ReadSachdev90} equations (\ref{rg-eqs})  estimate this critical value
as follows:
\begin{equation} 
\label{ncr} 
(n_{c}/N)_{\rm crit}=(1+1/\lambda)^{1/2}/(2\pi).
\end{equation}
On the isotropic ($\lambda=1$) square lattice, this yields
$(n_{c}/N)_{\rm cr}=(2\pi^{2})^{-1/2}\simeq 0.225$, which agrees qualitatively with the
value of $0.19$ obtained by a mean-field large-$N$ solution,\cite{ArovasAuerbach88} 
and with the numerical studies\cite{Assaad05,KawashimaTanabe07,Beach+09}  suggesting that 
the system has no N\'eel order for  $N/n_{c}\geq 5$. 

\begin{figure}[tb]
 \includegraphics[width=0.47\textwidth]{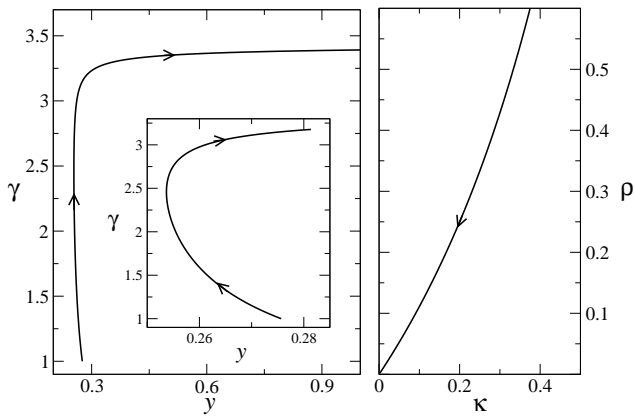}
 \caption{
\label{fig:type2}
A typical flow of the couplings in the disordered (VBS) phase in $d=2$.
The inset shows the ``U-turn'' behavior.}
\end{figure}

For nonzero $J'$, we have studied the equations (\ref{rg-eqs})
numerically for different values of the lattice anisotropy $\lambda$
and the $Sp(4)$ perturbation $J'$. 
In the two-dimensional case ($d=2$),
they exhibit two different characteristic flow patterns: In type I, $y$ flows to zero as
$l\to\infty$, while the other couplings flow to some constant values.
This type of flow corresponds to the N\'eel-ordered phase, and the Lorentz
invariance remains broken: there are two different velocities $u_{1}=\sqrt{\gamma}$
and $u_{2}=u_{1}\sqrt{(1+\rho)/(1-\kappa)}$. 

 \begin{figure}[tb]
 \includegraphics[width=0.4\textwidth]{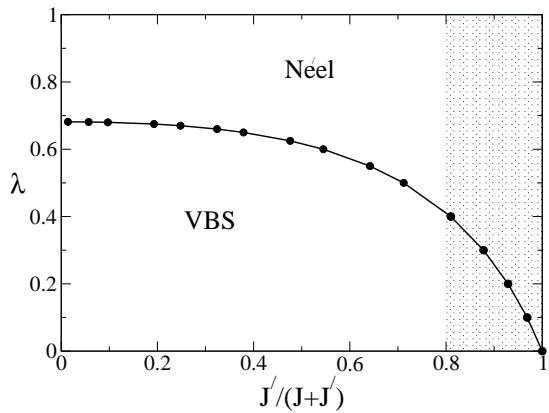}
 \caption{
\label{fig:kappa-lambda}
A schematic phase diagram of the $Sp(4)$ Heisenberg model
(\ref{ham-spin}) on a two-dimensional rectangular lattice in absence
of the quadratic Zeeman field, obtained from the analysis of the RG
equations (\ref{rg-eqs}). The shaded region close to the uniform
$SU(4)$ antiferromagnetic point $J=0$ indicates that the applicability of 
our effective theory in that domain is questionable.
}
\end{figure}

In type II, $y$ flows to
infinity at a certain scale $l=l_{0}$ as $y\sim 1/(l_{0}-l)$, while
$\gamma$ flows to a constant, and both $\rho$ and $\kappa$ flow to
zero as $\rho,\kappa\sim(l_{0}-l)^{1-3/2N}$. This behavior corresponds to a disordered phase, the
Lorentz invariance is  restored (the perturbation terms in
(\ref{Aeff}) flow to zero), so in the disordered phase the system is
again effectively described by
the $CP^{N-1}$ model, albeit  with a  renormalized velocity 
$u=\sqrt{\gamma}$ and the effective coupling
$g=y/(2C_{d=2})$. In this phase, the presence of a topological term induces
dimerization,\cite{ReadSachdev90} so this regime in fact corresponds
to a valence bond solid (VBS) state.

It is worth noting that when $y$ flows to infinity in $d=2$, it exhibits a
two-stage ``U-turn'' behavior as shown in  Fig.\ \ref{fig:type2}: at
the initial stage of the flow, up
to a certain scale $l=l^{*}$,  $g$ decreases and only later starts
growing till it explodes at $l=l_{0}$. The scale $l^{*}$ increases in
the vicinity of the phase transition line (see below). 
This behavior is reminiscent of
a double-scale behavior observed in the $\rm SO(3)$ model
\cite{RaoSen94} as well as in the  $CP^{N-1}$ model with a massive
gauge field. \cite{AzariaLecheminantMouhanna95}

If at $J'=0$ we are in the VBS phase (i.e., $n_{c}/N$ is below the
critical value (\ref{ncr})), then increasing $\kappa^{(0)}\propto J'$
beyond some threshold $\kappa_{c}$ leads to a transition to the N\'eel
phase.  On the anisotropic (rectangular) lattice, with the increasing
anisotropy (deviation of $\lambda$ from $1$) the transition point
$\kappa_{c}$ shifts towards higher values. The resulting phase diagram
is shown in Fig.\ \ref{fig:kappa-lambda}.  Thus, in two dimensions the
$Sp(N)$ perturbation terms in (\ref{Aeff}) are dangerously
irrelevant and can drive the phase transition between the disordered
(VBS) and N\'eel phase.

In one dimension, $y$ always flows to infinity
indicating that the
system is dimerized in the entire range of $0\leq\kappa<1$,
in line with the numerical results. \cite{Temo-spin32}
Curiously, in the close vicinity of the uniform $SU(N)$ point $\kappa=1$ ($J'\gg J$),
the coupling $y$ again exhibits the ``U-turn'' behavior as described
above, and the intermediate scale $l^{*}$ seems to diverge as
$J/J'\to 0$. This agrees with the fact the uniform $SU(N)$
antiferromagnet 
$\kappa=1$ is gapless in $d=1$ (it corresponds to the exactly solvable
Uimin-Lai-Sutherland (ULS) model\cite{Uimin70Lai74Sutherland75}).

%%%%%%%%%%%%%%%%%%%%%%%%%%%%%%%%%%%%%%%%%%%%%%%%%%%%%%%%%%%%%%%%%%%%%%%%%%%%%%%%%%%%%%%%%%%%%%%%%%%%%%%%%%%%%%%%%%%%%%%%
With the present approach, we are not able to detect any tendency
towards a
transition to the VBS phase with increasing $J'/J$ for the case of isotropic square lattice (line
$\lambda=1$). One has to keep in mind
that 
our  construction of smooth fields becomes
increasingly inadequate as
$J\to 0$; however, one  can argue that the theory still remains valid at the
energy scales of less than order $J$.
Several
 numerical results using exact diagonalization
on small 2d clusters, \cite{Bossche+00} series expansions,\cite{Zasinas+01} and density matrix
renormalization group (DMRG) on a ladder \cite{Chen+05} suggest that 
the uniform $SU(4)$ antiferromagnet ($J=0$, $\lambda=1$) is in a VBS phase with the
plaquette-type dimerization order. At the same time, theoretical studies advocate different scenarios for the 
 uniform $SU(N)$ 
antiferromagnetic point: in a recent work  based on the Majorana
fermion representation of spin-orbital operators,\cite{Ashvin09} the existence of a $Z_2$ spin-orbital
liquid state with emergent nodal fermions has been proposed for $N=4$;
other studies based on Schwinger-boson representations
\cite{Shen02,Toth+10} and exact diagonalization for the $SU(3)$ case\cite{Toth+10} suggest that at this point the ground state  has
the N\'eel-type $N$-sublattice order, which may be viewed as order at
the wavevector $(2\pi/N,2\pi/N)$.  The question about the correct ground
state around the point $\lambda=1$, $J=0$ is thus still open.
Further, our result shown in
Fig.\ \ref{fig:kappa-lambda} indicates that the VBS phase present at
small $\lambda$ has the
tendency to shrink with increasing $J'/J$. This makes plausible to
assume that, even if the point $\lambda=1$, $J=0$ is in the VBS state,
this phase should be different from the VBS phase at small
$\lambda$. Another argument in favor of this scenario is the
following: consider the point $J=0$, $\lambda=0$ which describes
uncoupled ULS chains. Each chain is gapless, with zero gap at wave vectors $k=2\pi m/N$,
$-N/2 < m\leq N/2$. Switching on weak interaction $\lambda$
between the chains may be expected to lead to an immediate ordering at
those wave vectors, while switching on weak $J$ leads to a VBS state.

%%%%%%%%%%%%%%%%%%%%%%%%%%%%%%%%%%%%%%%%%%%%%%%%%%%%%%%%%%%%%%%%%%%%%%%%%%%%
\section{The effect of quadratic Zeeman field in the \boldmath $Sp(4)$
  Heisenberg model}
\label{sec:Q}
%%%%%%%%%%%%%%%%%%%%%%%%%%%%%%%%%%%%%%%%%%%%%%%%%%%%%%%%%%%%%%%%%%%%%%%%%%%%%%%%

Let us now add the quadratic Zeeman
field term (\ref{A-Zeeman}) to the effective  model (\ref{Aeff}) with $N=4$. 
Now the first and the fourth field components become massive and can be
integrated out at once. We decompose the field into the 
background part $\vec{\varphi}$ and the ``fast'' massive part
$\vec{\chi}$ as follows:
\begin{eqnarray} 
\label{slow-fast-anis}
&& \vec{z}=\vec{\varphi}\sqrt{1-|\vec{\chi}|^{2}}+\vec{\chi},\nonumber\\
&& \vec{\varphi}=(0,\varphi_{1},\varphi_{2},0),\quad
\vec{\chi}=(\chi_{1},0,0\chi_{4}),
\end{eqnarray}
where
$\vec{\varphi}^{*}\cdot\vec{\varphi}=1$. One can straightforwardly
show that for the two-component unit complex vector field the following identity holds:
\begin{equation} 
\label{id-2} 
|\widetilde{\vec{\varphi}}\cdot
\partial_{\mu}\vec{\varphi}|^{2}=|\partial_{\mu}\vec{\varphi}|^{2}-|\vec{\varphi}^{*}\cdot
\partial_{\mu}\vec{\varphi}|^{2}\equiv |D_{\mu}\vec{\varphi}|^{2}.
\end{equation}
Thus, integrating out
$\vec{\chi}$, one obtains the familiar $CP^{1}$ model  that is equivalent to the $\rm O(3)$ nonlinear sigma
model (NLSM) and has been extensively used for a description of
Heisenberg spin systems. 
The topological term (\ref{Atop}) is retained.
The resulting action takes the form:
\begin{equation} 
\label{A-CP1} 
\mathcal{A}_{CP^{1}}[\vec{\varphi}]=\frac{\Lambda^{d-1}}{2}\int d^{d+1}x \Big\{
\frac{|{D}_{0}\vec{\varphi}|^{2}}{g_{1}^{*}}
+\frac{{D}_{k}\vec{z}|^{2} }{g_{2}^{*}}
\Big\}+\mathcal{A}_{\rm top}[\vec{\varphi}].
\end{equation}
The renormalized couplings $g_{1,2}^{*}$ are given by the formulas:
\begin{equation} 
\label{g-CP1} 
\frac{1}{g_{a}^{*}}= \frac{1}{g_{a}^{(0)}}+\frac{1}{\widetilde{g}_{a}^{(0)}}
-4g_{1}^{(0)}L_{d}(q_{0})
\Big(\frac{1}{g_{a}}+\frac{2}{\widetilde{g_{a}}}),\quad a=1,2,
\end{equation}
where: 
\begin{eqnarray} 
\label{Ld-Delta} 
&& L_{d}\left(\frac{\Delta^{2}}{\Lambda^{2}}\right)
=\frac{\Lambda^{1-d}}{(2\pi)^{d+1}}\int_{-\infty}^{\infty}dk_{0}\int_{|\vec{k}|=0}^{\Lambda}
\frac{ d^{d}\vec{k} }{k_{0}^{2}+\vec{k}^{2}+\Delta^{2}},\nonumber\\
&& q_{0}\equiv\frac{\Delta^{2}}{\Lambda^{2}}=g_{1}^{(0)}m_{0}^{2}=\frac{2q}{J}.
\end{eqnarray}
In one and two dimensions one has respectively:
\begin{eqnarray} 
\label{Ld-gD-d12}
L_{1}(x)&=&\frac{1}{2\pi}\ln\Big\{ 
\frac{1+\sqrt{1+x}}{\sqrt{x}}
\Big\},\nonumber\\
 L_{2}(x)&=&\frac{1}{4\pi}\left\{ 
\sqrt{1+x} -\sqrt{x}
\right\}.
\end{eqnarray}

%%%%%%%%%%%%%%%%%%%%
 \begin{figure}[tb]
 \includegraphics[width=0.4\textwidth]{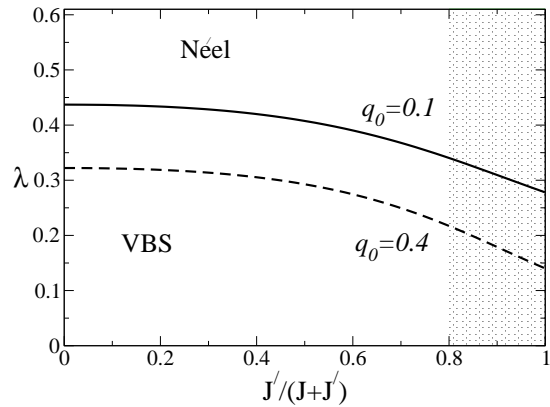}
 \caption{
\label{fig:kappa-lambda-anis}
A schematic phase diagram of the $Sp(4)$ Heisenberg model
(\ref{ham-spin}) on a two-dimensional rectangular lattice, in presence
of a finite quadratic Zeeman field $q=J q_{0}/2$. The transition line is
determined by Eq.\ (\ref{PT-anis}) with $g_{c}=\pi$; it moves towards
higher $\lambda$ with decreasing $q$ and  becomes flat at $q=0$. 
The shaded region close to the uniform
$SU(4)$ antiferromagnetic point $J=0$ indicates that the applicability of 
our effective theory in that domain cannot be trusted.
}
\end{figure}
%%%%%%%%%%%%%%

The $CP^{1}$ model with the topological phase angle $\theta=\pi$
has a disordering  transition into a gapped dimerized (VBS)
phase above a certain critical
value $g_{c}$ of the effective coupling $g_{\rm
  eff}=\sqrt{g_{1}^{*}g_{2}^{*}}$, both in one and two spatial
dimensions. For $g_{\rm eff}< g_{c}$, the model is gapless, is long-range ordered in $d=2$
and  has quasi-long-range order in $d=1$. Thus, the line of phase
transition between the N\'eel and VBS phases is given by:
\begin{equation} 
\label{PT-anis} 
g_{1}^{*}g_{2}^{*}=g_{c}^{2}
\end{equation}
Although the exact value of $g_{c}$ is not known, one may expect that
Eq.\ (\ref{PT-anis}) will qualitatively reproduce the transition
line. 

Fig.\ \ref{fig:kappa-lambda-anis} shows the result for the case of 
two dimensions, where  we have 
used $g_{c}=\pi$ (which is just the extrapolation of the large-$N$
result $g_{c}=\frac{2\pi}{N}$ down to the case $N=2$). One can see
that the curvature of the phase boundary agrees with the results of the
previous section obtained in absence of the field. Again, we cannot
see any tendency toward the VBS order at the uniform $SU(N)$
antiferromagnetic point on a square
lattice ($J=0$, $\lambda=1$). 

%%%%%%%%%%%%%%%%%%%
 \begin{figure}[tb]
 \includegraphics[width=0.42\textwidth]{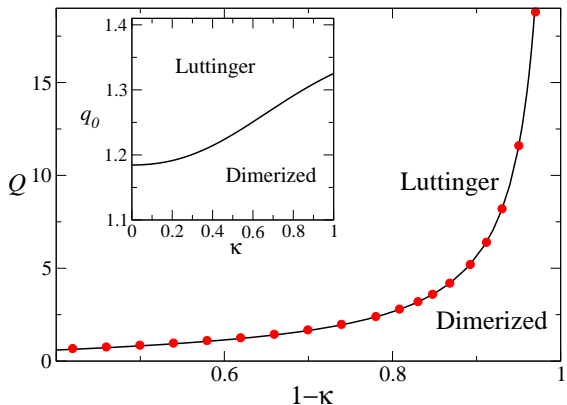}
 \caption{
\label{fig:anis-j1j2}
A schematic phase diagram of the $Sp(4)$ Heisenberg model
(\ref{ham-spin}) in one dimension, in presence of a finite quadratic
Zeeman field $q$.  The transition boundary (solid line) obtained from
the present $CP^{1}$ model approach, Eq.\ (\ref{PT-anis}) for $d=1$
with $g_{c}=2.1$, is compared to the result of
Ref.\ \onlinecite{Temo-spin32} (filled circles) obtained by  mapping
to an effective spin-$\frac{1}{2}$ zigzag chain. The parameters on the axes are given
by Eq.\ (\ref{Q-Temo}). The inset shows the same transition boundary
in the full range of $\kappa$ in terms of $q_{0}=2q/J$.  }
\end{figure}
%%%%%%%%%%%%%%%%%%

In the one-dimensional case, one can do better and extract the value
of $g_{c}$ from the comparison with the transition in an
antiferromagnetic spin-$\frac{1}{2}$ zigzag chain.
Rodriguez et al.\cite{Temo-spin32} have used a direct mapping of the
original fermionic model (\ref{ham-atom}) onto an effective
spin-$\frac{1}{2}$ chain with nearest- and next-nearest-neighbor
exchange couplings $j_{1}$ and $j_{2}$, respectively. The constants
$j_{1,2}$ were obtained as series in the perturbation parameter $1/Q$,
defined as follows: 
\begin{equation} 
\label{Q-Temo} 
Q=\frac{q(U_{2}+U_{0})}{2t^{2}}\equiv\frac{2q(1-\kappa)}{J\kappa(1-\kappa)},
\quad \kappa=\frac{J'}{J+J'}.
\end{equation}
Further, by using the value
$j_{2}/j_{1}\simeq 0.24$ for the transition point into the dimerized
phase, which is known from numerical
studies,\cite{OkamotoNomura92} in Ref.\ \onlinecite{Temo-spin32} an
estimate for the transition line in the $(1-\kappa,Q)$ plane has been
obtained that compared very well with the numerical results for the
original spin-$\frac{3}{2}$ fermionic model. 

In our approach, we can try and fix the value of $g_{c}$ (which is a
sole fitting parameter in our theory, for the entire line of
transition points in the $(\kappa,q)$ space) by comparing the output
of our Eq.\ (\ref{PT-anis}) to the results of Ref.\ \onlinecite{Temo-spin32}.
Fig.\ \ref{fig:anis-j1j2} shows the
transition line obtained from Eq.\ (\ref{PT-anis}) for $d=1$
with $g_{c}=8.5$ in comparison to the curve obtained in
Ref.\ \onlinecite{Temo-spin32}. One can see that the above value of
$g_{c}$ yields a good agreement close to the $SU(N)$
antiferromagnetic point $\kappa=0$. 
So, as a byproduct of our studies of the $Sp(4)$ Heisenberg model,
we obtain an independent estimate of the critical coupling value for the 1d $\rm
CP^{1}$ model  (or, alternatively, the $\rm O(3)$ NLSM)
at the topological angle $\theta=\pi$:
\begin{equation} 
\label{gc-CP1} 
 g_{c}^{CP^{1}}=\frac{1}{4}g_{c}^{NLSM} \simeq (2.1\pm 0.1).
\end{equation}

In the vicinity of the Sutherland point $J=0$ our description breaks
down; this can be seen already from the fact
that the transition line goes to a finite value of $q_{0}=2q/J$ at $J\to 0$
(see the inset of Fig.\ \ref{fig:anis-j1j2}), while  the main scale in
this limit is $J'\gg J$, so the critical value
of $q_{0}$ should diverge as $(1-\kappa)^{-1}$ in this limit.

%%%%%%%%%%%%%%%%%%%%%%%%%%%%%%%%%%%%%%%%%%%%%%%%%%%%%%%%%%%%%%%%%%%%%%%%%%%%%%%%
\section{Summary}
\label{sec:summary}

We have considered the model (\ref{ham-atom}) describing
spin-$\frac{3}{2}$ fermions in a spatially anisotropic optical lattice
shown in Fig.\ \ref{fig:lattice} at quarter filling
in the Mott limit of the on-site
repulsion constants $U_{0,2}$ being much larger than the hopping
amplitudes $t$. In this limit the charge degrees of freedom have a
large gap, and
the system can be mapped to the so-called $Sp(4)$ Heisenberg spin
model. 

We have studied its large-$N$ generalization, the $Sp(N)$ spin model,  with the help
of the field-theoretical approach constructed in the vicinity of the staggered $SU(N)$
antiferromagnetic point.  It is shown that the
effective field theory corresponds to the $Sp(N)$ extension of the
$CP^{N-1}$ model, with the Lorentz invariance generically broken
by the $Sp(N)$ terms that break the $SU(N)$ symmetry. For this effective field theory, we
have obtained the renormalization group equations to one-loop order
and have shown that although in the vicinity of the staggered $SU(N)$
antiferromagnetic point the $Sp(N)$ terms are seemingly
irrelevant, their
presence leads to a considerable renormalization of the $SU(N)$
part of the action, thus driving the transition between the
phase with a long-range N\'eel-type order and the magnetically
disordered valence bond solid (VBS) phase. 
We would like to note that solutions of the renormalization group
equations in the disordered phase exhibit a characteristic
double-scale behavior close to the N\'eel-VBS transition
boundary. Such a behavior is reminiscent to that encountered in other
frustrated models,\cite{RaoSen94,AzariaLecheminantMouhanna95} and is
also expected\cite{MurthySachdev90,MotrunichVishwanath04} in the framework of the deconfined
criticality conjecture.\cite{Senthil+04}

In addition to the $Sp(N)$ perturbation, we have also analyzed the
effect of the external magnetic field (quadratic Zeeman effect) and
established the qualitative form of the phase diagrams in one and two
spatial dimensions.  For the physical case $N=4$, at large values of
the quadratic Zeeman field the effective theory reduces to $CP^{1}$
model describing an isotropic Heisenberg antiferromagnet with a
pseudospin $\frac{1}{2}$.  Its ground state in two dimensions is
always in the long-range ordered (N\'eel) phase for the
pseudospin-$\frac{1}{2}$, and when the field is decreased, this state
either adiabatically evolves into the N\'eel phase of
spin-$\frac{3}{2}$ fermions (with the reduced N\'eel order\cite{KawashimaTanabe07}) or
undergoes a phase transition into the VBS state.  In one dimension,
there is a phase transition of the Berezinskii-Kosterlitz-Thouless
type that corresponds to the spontaneous dimerization transition in a
frustrated spin-$\frac{1}{2}$ chain with next-nearest neighbor
exchange.\cite{Temo-spin32} As a byproduct, by fitting our results to the available numerical
data,\cite{Temo-spin32} we have obtained an estimate for the critical
coupling of the $CP^{1}$ model with the $\theta=\pi$ topological term
in $(1+1)$ dimensions.

One last word of caution is in order:
since our effective theory is constructed
around the staggered $SU(N)$ antiferromagnetic point $J'=0$, it is not
expected to work well close to the other, uniform antiferromagnetic $SU(N)$ point
$J=0$. 
For that reason, we cannot exclude the presence of another phase
transition to the VBS phase in some
region around the  uniform $SU(N)$ point on the isotropic lattice
($J=0$, $\lambda=1$), as suggested in Ref.\ \onlinecite{QiXu08}.
Constructing an effective field theory describing the vicinity
of the uniform antiferromagnetic $SU(N)$ point remains a challenge for
the future work.

%%%%%%%%%%%%%%%%%%%%%%%%%%%%%%%%%%%%%%%%%%%%%%%%%%%%%%%%%%%%%%%%%%%%%%%%%%%%%%%%
\begin{acknowledgments}

A.K. gratefully acknowledges the hospitality of the
Institute for Theoretical Physics at the Leibniz University of
Hannover. This work has been supported by cluster of excellence QUEST (Center for Quantum
Engineering and Space-Time Research). T.V. acknowledges SCOPES Grant IZ73Z0-128058.

\end{acknowledgments}
%%%%%%%%%%%%%%%%%%%%%%%%%%%%%%%%%%%%%%%%%%%%%%%%%%%%%%%%%%%%%%%%%%%%%%%%%%%%%%%%

%\newpage
%%%%%%%%%%%%%%%%%%%%%%%%%%%%%%%%%%%%%%%%%%%%%%%%%%%%%%%%%%%%%%%%%%%%%%%%%%%%
% Specify following sections are appendices. Use \appendix* if there is
% only one appendix.
\appendix
%%%%%%%%%%%%%%%%%%%%%%%%%%%%%%%%%%%%%%%%%%%%%%%%%%%%%%%%%%%%%%%%%%%%%%%%%%%%%%%%
%%%%%%%%%%%%%%%%%%%%%%%%%%%%%%%%%%%%%%%%%%%%%%%%%%%%%%%%%%%%%%%%%%%%%%%%%%%%%%%%
%%%%%%%%%%%%%%%%%%%%%%%%%%%%%%%%%%%%%%%%%%%%%%%%%%%%%%%%%%%%%%%%%%%%%%%%%%%%%%%%
%%%%%%%%%%%%%%%%%%%%%%%%%%%%%%%%%%%%%%%%%%%%%%%%%%%%%%%%%%%%%%%%%%%%%%%%%%%%%%%%
%%%%%%%%%%%%%%%%%%%%%%%%%%%%%%%%%%%%%%%%%%%%%%%%%%%%%%%%%%%%%%%%%%%%%%%%%%%%%%%%
%%%%%%%%%%%%%%%%%%%%%%%%%%%%%%%%%%%%%%%%%%%%%%%%%%%%%%%%%%%%%%%%%%%%%%%%%%%%%%%%
%%%%%%%%%%%%%%%%%%%%%%%%%%%%%%%%%%%%%%%%%%%%%%%%%%%%%%%%%%%%%%%%%%%%%%%%%%%%%%%%
\section{\boldmath To the derivation of the effective  $Sp(N)$ field theory}
\label{app:micro}
%%%%%%%%%%%%%%%%%%%%%%%%%%%%%%%%%%%%%%%%%%%%%%%%%%%%%%%%%%%%%%%%%%%%%%%%%%%%%%%%

We integrate out the staggered field $\psi$ as well as the lagrange
multipliers $\mu_{1,2}$ from the effective action
$\mathcal{A}$ given by Eqs.\ (\ref{A0})-(\ref{Aint}). The equation of
motion for $\psi$ has the following form:
\begin{equation} 
\label{eqmot1} 
M \vec{\psi} +\partial_{\tau}\vec{z}+\mu_{1}\vec{z}+\mu_{2}\vec{\psi}=0,
\end{equation}
where the matrix $M$ is given by 
\begin{equation} 
\label{matM} 
M=c(1-\Omega+\varepsilon P),\quad
\Omega_{\alpha\beta}=z_{\alpha}z^{*}_{\beta},\quad
P_{\alpha\beta}=\widetilde{z}_{\alpha}^{*}\widetilde{z}_{\beta}.
\end{equation}
Here $c=4J(d-1+\lambda)$, $\varepsilon=J'/J$, and
$\widetilde{\vec{z}}$ is defined by (\ref{ztilde}).  From the equation
of motion for $z$ one can conclude that $\mu_{2}$ is of the order of a
second derivative of $\vec{z}$ and thus can be neglected.

Since we have assumed $|\vec{\psi}|\ll 1$, one can exploit the approximately
holding identities $P^{2}\simeq P$, $\Omega^{2}\simeq \Omega$, $\Omega
P\simeq P\Omega\simeq 0$, $M\vec{z}\simeq 0$,
$\Omega\widetilde{z}^{*}\simeq 0$, etc., and their derivates such as  
\begin{eqnarray} 
\label{idents} 
&& 
(1-\Omega-P)\vec{z}\simeq (1-\Omega-P)\widetilde{\vec{z}}^{*}\simeq 0,\nonumber\\
&& (1-\Omega-P)(1-\Omega+\varepsilon P)\simeq 1-\Omega-P.
\end{eqnarray}
Applying (\ref{idents}) to (\ref{eqmot1}), we get $\mu_{1}\simeq
-(\vec{z}^{*}\partial_{\tau}\vec{z})$, and 
\begin{equation} 
\label{eqpsi2} 
(1-\Omega-P)(c\vec{\psi}+\partial_{\tau}\vec{z})=0.
\end{equation}
From the last equation one concludes that
\begin{equation} 
\label{psi1} 
\vec{\psi}=(-\partial_{\tau}\vec{z}+\vec{y})/c, 
\end{equation}
where $\vec{y}$ satisfies $(1-\Omega-P)\vec{y}=0$.
Expanding  $\vec{y}$ in the system of mutually orthogonal vectors $(\vec{z},
\widetilde{\vec{z}}^{*}, \vec{e}_{a})$:
\[
\vec{y}=w_{1}\vec{z} +w_{2}\widetilde{\vec{z}}^{*}+\sum_{a=3}^{N} w_{a}\vec{e}_{a},
\]
and applying
$(1-\Omega-P)$ to $y$, one readily obtains
\begin{equation} 
\label{y} 
\vec{y}=w_{1}\vec{z}+w_{2}\widetilde{\vec{z}}^{*}. 
\end{equation}
Substituting the
above result for $\vec{y}$ back into (\ref{eqmot1}) yields 
\begin{equation} 
\label{w2} 
w_{2}\simeq
\frac{\varepsilon}{1+\varepsilon}(\widetilde{\vec{z}}\cdot\partial_{\tau}\vec{z}),
\end{equation}
and further, using the constraint $\vec{\psi}\cdot
\vec{z}^{*}+\vec{\psi}^{*}\cdot\vec{\psi}=0$, one obtains for the coefficient $w_{1}$:
\begin{equation} 
\label{w1} 
w_{1}+w_{1}^{*}=2\partial_{\tau}\vec{z}\cdot \vec{z}^{*}.
\end{equation}
Collecting Eqs.\ (\ref{psi1})-(\ref{w1}), one obtains the
resulting expression for $\psi$ :
\begin{equation} 
\label{psi2} 
\vec{\psi}=\frac{1}{c}\left(-\partial_{\tau}\vec{z}+\frac{\varepsilon}{1+\varepsilon}
(\widetilde{\vec{z}}\cdot\partial_{\tau}\vec{z})\widetilde{\vec{z}}^{*} +w_{1}\vec{z}\right).
\end{equation}
Substituting the above expression back into the action $\mathcal{A}[z,\psi]$, one
finally obtains the effective action in the form (\ref{Aeff}). 

%%%%%%%%%%%%%%%%%%%%%%%%%%%%%%%%%%%%%%%%%%%%%%%%%%%%%%%%%%%%%%%%%%%%%%%%%%%%%%%%
\section{\boldmath Linear excitation analysis for the lattice $Sp(N)$
  model }
\label{app:sw}
%%%%%%%%%%%%%%%%%%%%%%%%%%%%%%%%%%%%%%%%%%%%%%%%%%%%%%%%%%%%%%%%%%%%%%%%%%%%%%%%

Here we provide the classical linear excitation analysis for the lattice
Lagrangian (\ref{Llat}). Let us
choose the classical ground state as $\vec{z}=(1,0,\ldots,0)$, then 
small deviations from this ground state can be written as
$\vec{z}_{n}=(\sqrt{1-|\vec{\varphi}_{n}|^{2}}, \vec{\varphi}_{n})$,
where $\vec{\varphi}_{n}=(\varphi_{1n},\ldots,\varphi_{N-1,n})$ is a
$(N-1)$-component complex vector. Expanding in $\vec{\varphi}$, and keeping only up to quadratic terms,
 we obtain the Lagrangian:
\begin{eqnarray} 
\label{Lsw}
\mathcal{L}&=&-i\sum_{n}\eta_{n}\vec{\varphi}_{n}^{*}\cdot\partial_{t}\vec{\varphi}_{n}
-Z\sum_{n}(J|\vec{\varphi}_{n}|^{2} +J'|\varphi_{N-1,n}|^{2})
\nonumber\\
&+&\frac{1}{2}\sum_{\langle n n'\rangle}\big\{ 
J(\vec{\varphi}_{n}^{*}\cdot\vec{\varphi}_{n'}+\varphi_{N-1,n}^{*}\varphi_{N-1,n'}
+ \mbox{c.c.})
\big\},
\end{eqnarray}
where $Z=2d$ is the lattice coordination number, and for the sake of
clarity we have switched back to real time $t$ and set the lattice to be
spatially isotropic ($\lambda=1$). After the standard Fourier
transform, the equations of motion are obtained as
$i\partial_{t}\varphi_{a}(k)+F_{a}(k)\varphi_{a}(k)=0$, where the
functions $F_{a}$ are given by:
\begin{eqnarray} 
\label{Fa} 
F_{a}(k)&=&2J\sum_{\mu=1}^{d}(1+\cos k_{\mu}),\quad
a=1,\ldots,N-2\nonumber\\
F_{N-1}(k)&=&2(J+J')\sum_{\mu=1}^{d}(1+\cos k_{\mu}).
\end{eqnarray}
The dispersions $\omega_{a}(k)$ of linear modes (``spin waves'') are thus determined
simply by the relation $\omega_{a}^{2}(k)=F_{a}(k)F_{a}(k+\pi)$, which
in the limit $\vec{k}\to 0$ yields the spin wave velocities:
\begin{eqnarray} 
\label{vSW}
&& v_{a}=2J\sqrt{d} ,\quad a=1,\ldots,N-2,\nonumber\\
&& v_{N-1}=2(J+J')\sqrt{d}.
\end{eqnarray}
Those velocities, obtained by a spin-wave-type lattice calculation,
perfectly agree with the velocities obtained from our effective
continuum action (\ref{Aeff}).

As a side remark, it is worthwhile to note that in presence of the quadratic Zeeman
field $q$ (see (\ref{q-Zeeman})) the spin wave velocities 
do not change with the increase of $q$, counterintuitively to the
common knowledge that the spin wave velocity is linearly proportional
to the spin magnitude $S$ (and $S$ effectively decreases from
$\frac{3}{2}$ at $q=0$ to $S=\frac{1}{2}$ at $q=\infty$, for the
physical case $N=4$). For $N=4$, the effect of
QZE is to make two out of three spin wave modes massive, but
it does not touch the velocities (which, for gapped modes, take the
meaning of limiting velocities).
On the other hand, when one changes the spin-$2$ channel interaction $U_{2}$ from
$+\infty$ to $U_{0}$ (which corresponds to the path from the staggered to the uniform
antiferromagnetic $SU(4)$ points), velocities $v_{1,2}$ decrease and tend to zero
as $U_{2}\to U_{0}$, while the remaining velocity $v_{3}$ increases. Particularly, $v_{3}$ is twice as large at the
uniform $SU(4)$ point as it is at the staggered $SU(4)$ one,
$v_{3}(U_{2}=U_{0})/v_{3}(U_{2}=\infty)=2$. 
Physically, softening of $v_{1,2}$  
reflects the increase of frustration on the way from the staggered to
the uniform antiferromagnetic $SU(4)$ point.

%%%%%%%%%%%%%%%%%%%%%%%%%%%%%%%%%%%%%%%%%%%%%%%%%%%%%%%%%%%%%%%%%%%%%%%%%%%%%%%%
\section{\boldmath RG equations for the $Sp(N)$ model}
\label{app:rg}
%%%%%%%%%%%%%%%%%%%%%%%%%%%%%%%%%%%%%%%%%%%%%%%%%%%%%%%%%%%%%%%%%%%%%%%%%%%%%%%%

We derive here the RG equations for our $Sp(N)$ effective action without Lorentz invariance
(\ref{Aeff}), using the Polyakov background field method.\cite{Polyakov75} 
We start by splitting the fields $z_{\alpha}$ and $\mathcal{A}_{\mu}$
into the background (``slow'') fields $\varphi_{\alpha}$, $A_{\mu}$,
  and the
fluctuation (``fast'') parts $\chi_{a}$, $a_{\mu}$:
\begin{eqnarray} 
\label{slow-fast} 
\vec{z}&=&\vec{\varphi}\sqrt{1-|\vec{\chi}|^{2}}+\sum_{a=1}^{N-1}\chi_{a}\vec{e}_{a},\nonumber\\
\mathcal{A}_{\mu}&=&A_{\mu}+a_{\mu},
\end{eqnarray}
where $\{\vec{\varphi},\vec{e}_{a} \}$ form a set of
mutually orthogonal complex unit vectors. Since the ``tilded'' slow
field $\widetilde{\vec{\varphi}}=\mathcal{J}\vec{\varphi}$ satisfies the condition
$\widetilde{\vec{\varphi}}\cdot \vec{\varphi}=0$, we can expand it as
\begin{equation} 
\label{slow-tilded}
\widetilde{\vec{\varphi}}=\sum_{a}\widetilde{\varphi}_{a} \vec{e}_{a}^{*}. 
\end{equation}
We will not need any explicit expansion of  the ``tilded'' fast field
$\widetilde{\vec{\chi}}$, because we will be able to avoid its presence by  using
the identities $\vec{x}\cdot
\widetilde{\vec{y}}=-(\widetilde{\vec{x}}\cdot \vec{y})$, 
$\widetilde{\vec{x}}\cdot\widetilde{\vec{x}}=\vec{x}\cdot\vec{x}$.
We will use the notation
$\mathcal{D}_{\mu}=\partial_{\mu}-i\mathcal{A}_{\mu}$ and $D_{\mu}=\partial_{\mu}-iA_{\mu}$.
 The derivatives of $\{
\vec{e}_{a}(x)\}$ can be written in the form
\begin{eqnarray} 
\label{e-der} 
\partial_{\mu}\vec{e}_{a}&=&B^{ab}_{\mu}\vec{e}_{b}+B^{a0}_{\mu}\vec{\varphi},\nonumber\\
\partial_{\mu}\vec{\varphi}&=&B^{0a}_{\mu}\vec{e}_{a}+B^{00}_{\mu}\vec{\varphi},
\end{eqnarray}
where
$B^{\alpha\beta}_{\mu}=-(B_{\mu}^{\beta\alpha})^{*}=\vec{e}_{\beta}\cdot\vec{e}_{\alpha}$,
$\vec{e}_{0}\equiv \vec{\varphi}$. The quantity
$B^{00}_{\mu}=\vec{\varphi}^{*}\cdot \partial_{\mu}\vec{\varphi}$ can
be eventually identified with $A_{\mu}$.

There is a substantial freedom in the choice of the local basis $\{
\vec{e}_{a}(x)\}$, which one can use in order to eliminate 
$B^{ab}_{\mu}$ (but not $B^{a0}_{\mu}$). Indeed, under a local unitary rotation $\vec{e}_{a}\mapsto
U_{ab}\vec{e}_{b}$ the $(N-1)\times(N-1)$ matrix $\bm{B}_{\mu}=\{B^{ab}_{\mu}\}$
transforms as $\bm{B}_{\mu}\mapsto (\partial_{\mu}
\bm{U})\bm{U}^{\dag}+\bm{U} \bm{B}_{\mu}\bm{U}^{\dag} $. Thus, to
eliminate $\bm{B}_{\mu}$, one has to solve the equation
$\bm{B}_{\mu}=(\partial_{\mu}\bm{U}^{\dag})U$. Comparing that to
(\ref{e-der}), it is easy to see that setting
$U_{ab}=(\vec{e}^{*}_{b})_{a}$ does the desired job.
Finally, multiplying the rotation matrix by a phase factor,
$U\mapsto U \exp\Big\{-i\int^{x} A_{\mu}(x')dx'_{\mu}\Big\}$, we can
eliminate the $\rm U(1)$ gauge field $A_{\mu}$ from the expression for
$D_{\mu}\vec{\chi}$ as well, so that one effectively replaces
$D_{\mu}\vec{\chi}$ by $\partial_{\mu}\chi_{a}\vec{e}_{a}+\chi_{a}B^{a0}_{\mu}\vec{\varphi}$.

We substitute the ansatz (\ref{slow-fast}) into the action
(\ref{Aeff}), and make use of the trick described above to simplify
$D_{\mu}\vec{\chi}$.
The ``fast'' component of the gauge field $a_{\mu}$ enters the action
in a quadratic way; integrating it out yields:
\begin{equation} 
\label{amu} 
a_{\mu}=i(\vec{\chi}\cdot(D_{\mu}\varphi)^{*}-\vec{\chi}^{*}\cdot D_{\mu}\vec{\varphi}).
\end{equation}
Plugging this expression back into the action,
one
obtains after some algebra the new effective action in the form
$\mathcal{A}_{\rm eff}[\vec{z}] =\mathcal{A}_{\rm eff}[\vec{\varphi}] 
+\mathcal{A}_{\rm int}$, with
\begin{eqnarray} 
\label{A-phi-chi}
\mathcal{A}_{\rm
  int}&=&\frac{\Lambda^{d-1}}{2}\int
d^{d+1}x \Big\{
(\partial_{\mu}\chi_{a}^{*})(\partial_{\mu}\chi_{b})\Big(\frac{\delta_{ab}}{g_{\mu}}
+\frac{\widetilde{\varphi}_{a}^{*}\widetilde{\varphi}_{b}}{\widetilde{g_{\mu}}}\Big) \nonumber\\
&+&\chi_{a}^{*}\chi_{b}\Big[ 
-\frac{\delta_{ab}}{g_{\mu}}|D_{\mu}\vec{\varphi}|^{2}
-\frac{2\delta_{ab}}{\widetilde{g_{\mu}}}|\widetilde{\vec{\varphi}}\cdot
D_{\mu}\vec{\varphi}|^{2}\nonumber\\
&-&\frac{1}{g_{\mu}}B^{0a}_{\mu}\big(B^{0b}_{\mu}\big)^{*} 
+\frac{1}{\widetilde{g_{\mu}}}
\big(D_{\mu}\widetilde{\varphi}_{a}\big)^{*}\big(D_{\mu}\widetilde{\varphi}_{b}\big)
\Big]\Big\}.
\end{eqnarray}
Here for the sake of brevity we have introduced the following notation: 
\begin{equation} 
\label{gmu} 
g_{\mu}=g_{2}+(g_{1}-g_{2})\delta_{0\mu},\quad
\widetilde{g_{\mu}}=\widetilde{g_{2}}+(\widetilde{g_{1}}-\widetilde{g_{2}})\delta_{0\mu},
\end{equation}
so that
$g_{\mu}=g_{1}$, $\widetilde{g_{\mu}}=\widetilde{g_{1}}$ for
$\mu=0$, and $g_{\mu}=g_{2}$, $\widetilde{g_{\mu}}=\widetilde{g_{2}}$
for $\mu\not=0$. 

Doing the final step of integrating out the fluctuations field
$\vec{\chi}$, it is convenient to make use of the following formula:
for a  matrix  of the form
\begin{equation} 
\label{matr} 
\widehat{M}=x \widehat{\openone} +y\widehat{P},\quad 
P_{ab}=\widetilde{\varphi}_{a}^{*}\widetilde{\varphi}_{b},
\end{equation}
where $\widetilde{\varphi}_{a}^{*}\widetilde{\varphi}_{a}=1$,  
its inverse can be explicitly written down as:
\begin{equation} 
\label{inv-matr} 
\widehat{M}^{-1}=\frac{1}{x} \widehat{\openone}
-\frac{y}{x(x+y)}\widehat{P}.
\end{equation}
With the help of this identity, the ``fast'' field $\vec{\chi}$,
containing the Fourier components with momenta in the interval
$[\Lambda,\Lambda(1+dl)]$, can be easily integrated out. A typical
integral over the momentum $(k_{0},\vec{k})$ has the form:
\begin{equation} 
\label{k-int}
\frac{1}{(2\pi)^{d+1}}\int_{-\infty}^{\infty}dk_{0}\int_{|\vec{k}|=\Lambda(1-dl)}^{\Lambda}
\frac{ d^{d}\vec{k} }{k_{0}^{2}+\gamma\vec{k}^{2}} = \Lambda^{d-1}
 \frac{C_{d}}{\sqrt{\gamma}} dl, 
\end{equation}
where we have denoted
 $C_{d}=\pi S_{d}/(2\pi)^{d+1}$ and
$S_{d}=2\pi^{d/2}/\Gamma(d/2)$ is the surface of a $d$-dimensional sphere.
The correction $\Delta \mathcal{A}$ to the action $\mathcal{A}_{\rm eff}[\vec{\varphi}]$,
coming from the fluctuation, takes the form:
\begin{eqnarray} 
\label{Acorr} 
\Delta \mathcal{A}&=&C_{d}\Lambda^{d-1} dl \int d^{d+1}x \Big\{
|D_{\mu}\vec{\varphi}|^{2} \\
&\times& \left[ \Big( \frac{g_{1}}{\widetilde{g}_{\mu}}
  -N \frac{g_{1}}{g_{\mu}} \Big)\frac{1}{\sqrt{\gamma}}
-\Big(\frac{g_{1}'}{\sqrt{\gamma'}}
   -\frac{g_{1}}{\sqrt{\gamma}}\Big)\frac{1}{g_{\mu}} \right]\nonumber\\
&+&|\widetilde{\vec{\varphi}}D_{\mu}\vec{\varphi}|^{2} \left[ 
\frac{2(1-N)g_{1}}{\widetilde{g}_{\mu}\sqrt{\gamma}}\right. \nonumber\\
&-&\left.\Big(\frac{2}{\widetilde{g}_{\mu}} +\frac{1}{g_{\mu}}\Big)
\Big(\frac{g_{1}'}{\sqrt{\gamma'}}-\frac{g_{1}}{\sqrt{\gamma}} \Big)
\right]\nonumber,
\end{eqnarray}
where we have introduced the shorthand notation
\begin{equation} 
\label{g-gamma} 
\gamma=\frac{g_{1}}{g_{2}},\quad 
g_{1}'=\frac{g_{1}\widetilde{g}_{1}}{g_{1}+\widetilde{g}_{1}},\quad
\gamma'=g_{1}'\frac{g_{2}+\widetilde{g}_{2}}{g_{2}\widetilde{g}_{2}},
\end{equation}
and have
again used the notation (\ref{gmu}). From (\ref{Acorr}), the RG
equations for the couplings are
readily obtained in the form
\[
df/dl=(1-d)f-2C_{d}\beta_{f}(f),
\]
where $f\in\{g_{1},g_{2}, \widetilde{g_{1}}, \widetilde{g_{2}} \}$, and the beta functions
are given by: 
\begin{eqnarray} 
\label{RG-raw}
\beta_{g_{1}}&=& g_{1}^{2} \Big\{  
\frac{1}{\sqrt{\gamma}}\Big(\frac{g_{1}}{\widetilde{g_{1}}} -N+1 \Big)
-\frac{g_{1}'}{g_{1}\sqrt{\gamma'}}
\Big\}, \\
\beta_{g_{2}}&=& g_{2}^{2} \Big\{  
\frac{1}{\sqrt{\gamma}}\Big(\frac{g_{1}}{\widetilde{g_{2}}} -(N-1)\frac{g_{1}}{g_{2}} \Big)
-\frac{g_{1}'}{g_{2}\sqrt{\gamma'}}
\Big\}, \nonumber\\
\beta_{\widetilde{g_{1}}}&=& \widetilde{g_{1}}^{2} \Big\{
\frac{1}{\sqrt{\gamma}}\Big( 1-(N-2)\frac{2g_{1}}{\widetilde{g_{1}}} \Big)
-\frac{1}{\sqrt{\gamma'}}\Big( \frac{g_{1}'}{g_{1}}+ \frac{2g_{1}'}{\widetilde{g_{1}}}\Big)
\Big\},\nonumber\\
\beta_{\widetilde{g_{2}}}&=& \widetilde{g_{2}}^{2} \Big\{
\frac{1}{\sqrt{\gamma}}\Big( \frac{g_{1}}{g_{2}} -(N-2)\frac{2g_{1}}{\widetilde{g_{2}}} \Big)
+\frac{1}{\sqrt{\gamma'}}\Big( \frac{g_{1}'}{g_{2}}+ \frac{2g_{1}'}{\widetilde{g_{2}}}\Big)
\Big\}.\nonumber
\end{eqnarray}
Rewriting the above system in variables (\ref{rg-pars}),
one finally obtains the RG equations in the form (\ref{rg-eqs}).

%%%%%%%%%%%%%%%%%%%%%%%%%%%%%%%%%%%%%%%%%%%%%%%%%%%%%%%%%%%%%%%%%%%%%%%%%%%%%%%%
% Create the reference section using BibTeX:
%\bibliography{basename of .bib file}
%%%%%%%%%%%%%%%%%%%%%%%%%%%%%%%%%%%%%%%%%%%%%%%%%%%%%%%%%%%%%%%%%%%%%%%%%%%%%%%%

\end{document}